\documentclass[letter]{aa}

\usepackage{graphicx}
\usepackage{natbib}
\usepackage{txfonts}
\begin{document}

   \title{Variability of the near-infrared extinction curve towards the Galactic centre}

   \author{F. Nogueras-Lara
          \inst{1}
          \and
           R. Sch\"odel 
          \inst{1}
          \and
           F. Najarro
          \inst{2}
         \and
          A. T. Gallego-Calvente
          \inst{1}
          \and
           E. Gallego-Cano
          \inst{1,3}
           \and
          B. Shahzamanian
          \inst{1}
         \and
          N. Neumayer
          \inst{4}
}

   \institute{
    Instituto de Astrof\'isica de Andaluc\'ia (CSIC),
     Glorieta de la Astronom\'ia s/n, 18008 Granada, Spain
              \email{fnoguer@iaa.es}
         \and
     Centro de Astrobiolog\'ia (CSIC/INTA), ctra. de Ajalvir km. 4, 28850 Torrej\'on de Ardoz, Madrid, Spain
         \and
     Centro Astron\'omico Hispano-Alem\'an (CSIC-MPG), Observatorio Astron\'omico de Calar Alto, Sierra de los Filabres, 04550, G\'ergal, Almer\'ia, Spain     
       \and
     Max-Planck Institute for Astronomy, K\"onigstuhl 17, 69117 Heidelberg, Germany
      }

   \date{}

% \abstract{}{}{}{}{} 
% 5 {} token are mandatory
 
  \abstract
   {Due to the extreme extinction towards the Galactic centre ($A_{V} \sim 30 $ mag), its stellar population is mainly studied in the near-infrared (NIR) regime. Therefore, a proper analysis of the NIR extinction curve is necessary to fully characterise the stellar structure and population of the inner part of the galaxy.}
   {We   studied  the dependence of the extinction index ($\alpha_\lambda$) in the NIR on the line of sight, wavelength, and extinction.} 
    {We used the GALACTICNUCLEUS imaging survey, a high angular resolution catalogue ($0.2''$) for the inner part of the Galaxy in $JHK_s$, and studied the spatial variation in the extinction index. We also applied two independent methods based on red clump stars to compute the extinction index between different bands and its variation with   wavelength.}
   {We did not detect any significant line-of-sight or extinction variation in $\alpha$ within the studied region in the nuclear stellar disc. The extinction index between $JH$ and $HK_s$ differs by $0.19 \pm 0.05$. We obtained mean values for the extinction indices $\alpha_{JH} = 2.43\pm0.03$ and $\alpha_{HK_s} = 2.23\pm0.03$. The dependence of the extinction index on the wavelength could explain the differences obtained for $\alpha_\lambda$ in the literature since it was assumed constant for the NIR regime.}

   {}


   \maketitle
%
%________________________________________________________________

\section{Introduction}

The Galactic centre (GC) is a crucial astrophysical laboratory since it is the closest galactic nucleus and the only one where we can resolve individual stars down to milliparsec scales. Nevertheless, very little is known about its structure and stellar population,     due to the strong crowding and the large interstellar extinction \citep[$A_V\gtrsim30$\,mag,
$A_{K_{s}}\gtrsim2.5$\,mag, e.g.][]{Scoville:2003la,Nishiyama:2008qa,Fritz:2011fk,Schodel:2010fk}. Therefore, a proper characterisation of the near-infrared (NIR) extinction law is fundamental to better understand the GC. 

It is generally accepted that the extinction curve in the NIR can be approximated by a power law \citep[e.g.][]{Nishiyama:2008qa,Fritz:2011fk} of the form $A_\lambda \propto \lambda^{-\alpha}$, where $\lambda$ and $\alpha$ are the wavelength and the extinction index, respectively. However, the value of the extinction index has changed significantly in recent decades from values of $\sim 1.5$ \citep[e.g.][]{Rieke:1985fq,Draine:1989eq} to $\alpha>2.0$ or even $\sim 2.5$  \citep[e.g.][]{Nishiyama:2006tx,Stead:2009uq,Gosling:2009kl,Schodel:2010fk,Fritz:2011fk,Alonso-Garcia:2017aa,Nogueras-Lara:2018aa}. In addition to this discrepancy, some evidence of a possible variation in the extinction index between the NIR bands $JH$ and $HK_s$ has been reported recently \citep{Nogueras-Lara:2018aa,Hosek:2018aa}. These different values can lead us to generate an incorrect picture of the inner structure of the galaxy. Namely, a small change in $\alpha$ ($\sim 10-15 \%$) can result in a change in absolute extinction of    $\sim 0.3$ mag, which corresponds to a bias in the estimation of distances, based on the distance modulus, of  $\sim 1$ kpc \citep{Matsunaga:2016aa} at the GC distance ($\sim 8$ kpc). The situation is even more complicated when inferring the stellar type of a star using NIR photometry, where a small variation correcting the extinction  completely changes the type of a star \citep[e.g. Figs. 33 and 34][]{Nogueras-Lara:2018aa}.

In this letter we characterise the extinction curve in the NIR bands $JHK_s$ towards the nuclear bulge (NB)  using the GALACTICNUCLEUS survey \citep{Nogueras-Lara:2018aa,Nogueras-Lara:2019aa} and two independent methods based on red clump (RC) stars \citep[e.g.][]{Girardi:2016fk}.

%In this letter we firstly characterise the extinction curve in the NIR bands $JHK_s$ towards the nuclear bulge (NB). We found that the extinction index does not vary with the line of sight within the studied region. Then, we applied two independent methods and find that the extinction index between $JH$ and $HK_s$ differs by $\sim 0.2$. This contradicts the accepted single power-law paradigm for the NIR. 

\section{Data}

We used for this study the GALACTICNUCLEUS survey \citep{Nogueras-Lara:2018aa,Nogueras-Lara:2019aa}. This is a $JHK_s$ NIR photometric survey carried out with the  HAWK-I camera \citep{Kissler-Patig:2008uq} located at the ESO VLT unit telescope 4. This survey uses the speckle holography technique described by \citet{Schodel:2013fk} to reach a high angular resolution of $0.2 ''$. The photometry and astrometry are obtained by means of point spread function (PSF) fitting using the {\emph StarFinder} software package \citep{Diolaiti:2000qo}. The catalogue reaches $5\,\sigma$   detection limits of approximately $J=22$, $H=21$, and $K_{s}=20$\,mag. The photometric uncertainties are less than $0.05$ mag at $J\lesssim20$, $H\lesssim17$, and $K_{s}\lesssim16$\,mag. The zero point (ZP) is calibrated using the SIRIUS/IRSF GC survey \citep[e.g.][]{Nagayama:2003fk,Nishiyama:2006tx} and its associated uncertainty is $\sim$ 0.036 mag in all three bands.

\begin{figure*}
   \centering
   \includegraphics[scale=0.33]{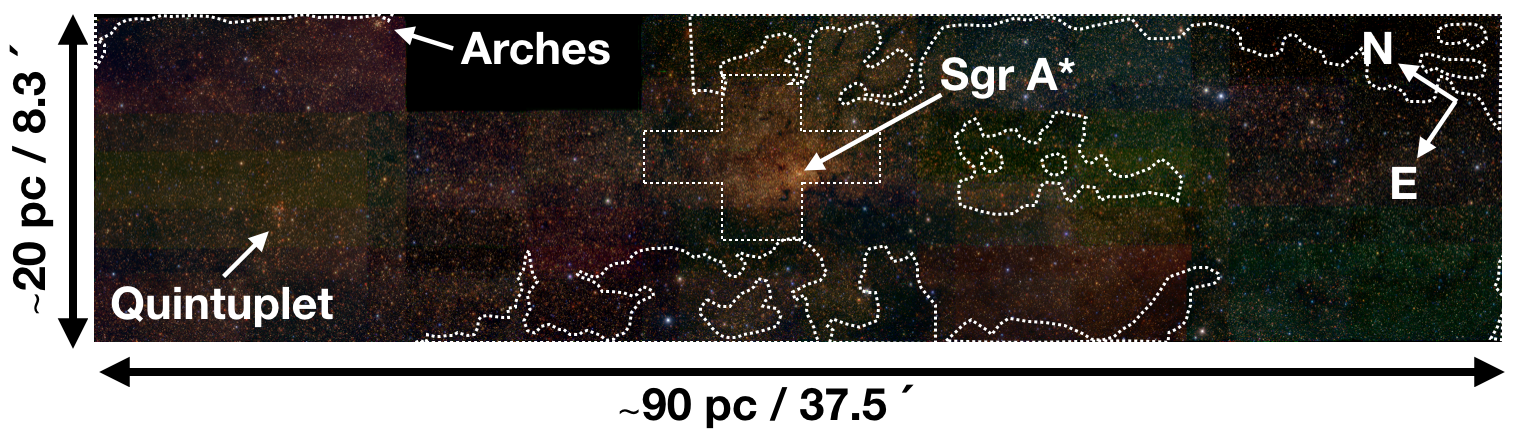}
   \caption{Image of the studied region produced combining the $K_s$, $H$, and $J$ bands in red, green, and blue, respectively. Sagittarius A* and the Arches and Quintuplet clusters are indicated by arrows. The black rectangle near the Arches cluster corresponds to a field with incomplete data. The white dashed contours indicate regions dominated by dark clouds. The cross-shaped region corresponds to a low completeness region due to crowding (the nuclear star cluster, NSC).}
   
   \label{RGB}
    \end{figure*}

In the study presented in this letter, we used the $J$, $H,$ and $K_s$ photometry of 14 different fields of the survey that cover a rectangular region of 90 pc $\times$ 20 pc centred on Sgr\,A* and corresponding to the central part of the NB \citep{Nogueras-Lara:2019ab}, as shown in Fig. \ref{RGB}.

\section{Colour-magnitude diagrams}

\begin{figure}
   \centering
   \includegraphics[width=\columnwidth]{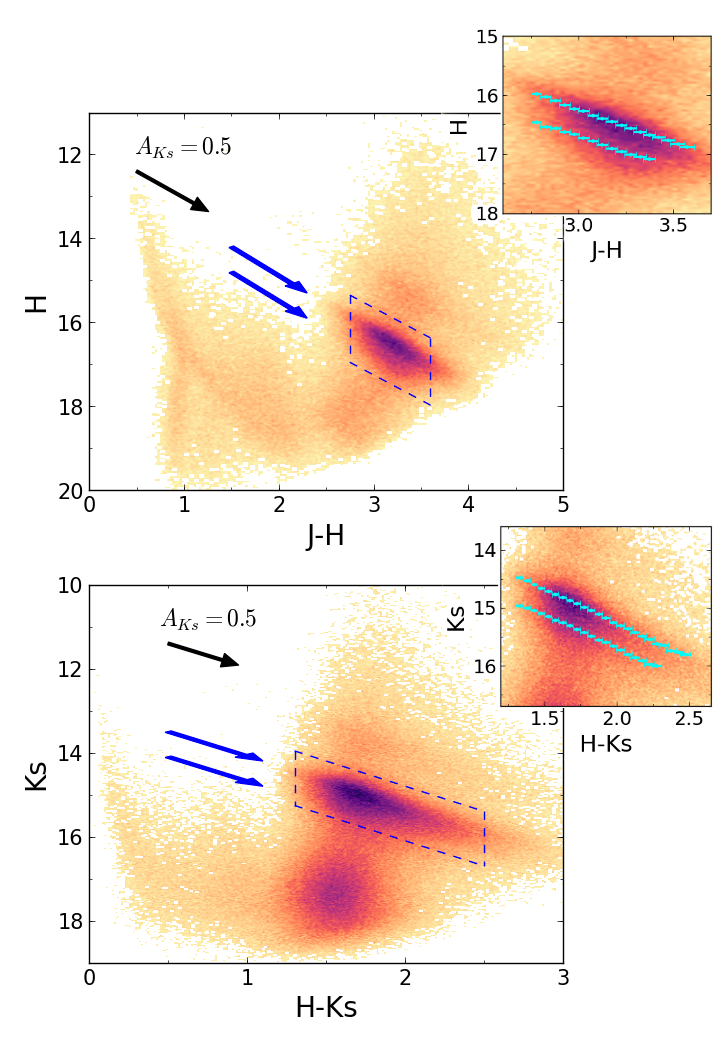}
   \caption{Colour-magnitude diagrams $H$ vs. $J-H$ (upper panel) and $K_s$ vs. $H-K_s$ (lower panel). The RC is marked by the blue dashed parallelograms. The two blue arrows show a double feature in the RC. The black arrow depicts the reddening vector with an extinction $A_{K_s} = 0.5$ mag \citep[computed using $\alpha=2.30\pm0.08$,][]{Nogueras-Lara:2018aa}. The insets show the RC region with the two features obtained applying GMM and their uncertainties in cyan (see main text).}
   
   \label{CMD}
    \end{figure}

Figure \ref{CMD} depicts the colour-magnitude diagrams (CMD) $H$ versus $J-H$ and $K_s$ versus $H-K_s$. The highly populated region located at $J-H>2.5$ and $H-K_s>1.3$ corresponds to the GC stellar population, whereas stars at $J-H<2.5$ and $H-K_s<1.3$ trace foreground stars probably belonging to three spiral arms \citep{Nogueras-Lara:2018aa}. The high density regions within the blue dashed trapezoids show the GC RC feature following the reddening vector due to differential extinction. We clearly distinguish a bright and a faint RC, which trace an old stellar population ($\gtrsim 8$ Gyr) and stars formed in a younger star formation burst ($\sim 1$ Gyr), respectively (Nogueras-Lara et al., submitted). 
%The feature at $H-K_s \sim 1.5$ and $K_s\sim 17$ is formed by ascending giant branch and post-main sequence stars.

\section{Extinction index analysis}

We analysed the line-of-sight variability and the variation as a function of the wavelength of the extinction index.

%To study the extinction curve, we assumed that it can be described by a power law of the form $A_\lambda \propto \lambda^{-\alpha}$ \citep[e.g.][]{Nishiyama:2008qa,Fritz:2011fk}, where $A_\lambda$ is the extinction at a given wavelength ($\lambda$) and $\alpha$ is the extinction index. 

\subsection{Spatial variability of the extinction index}
\label{grid_method}

\begin{figure}
   \centering
   \includegraphics[width=\columnwidth]{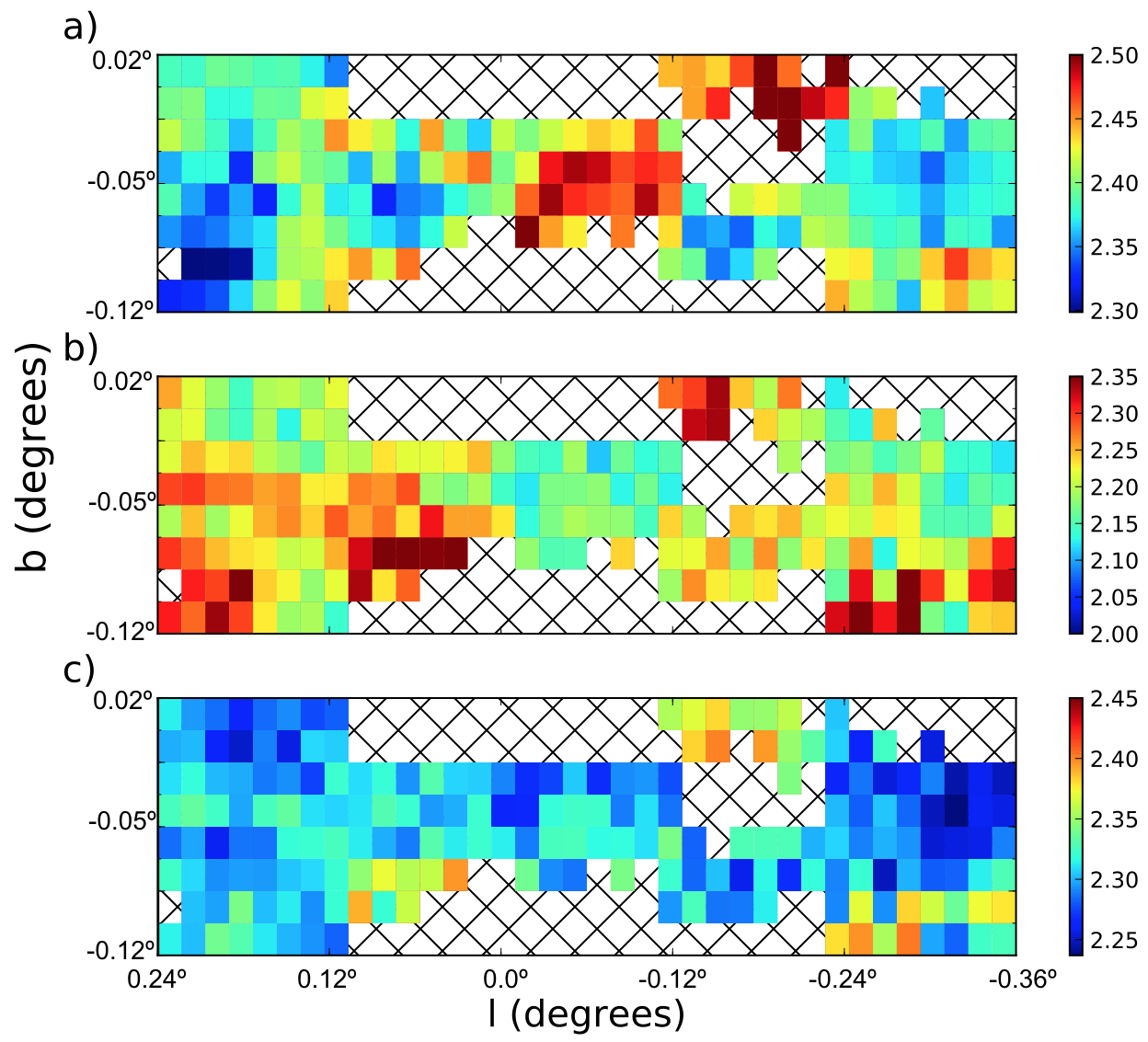}
   \caption{Extinction-index maps: a) $JH$-map; b) $HK_s$-map; c) $JK_s$-map. Cross-shaped pixels indicate that there are not enough stars for a reliable estimate.}
   
   \label{maps}
    \end{figure}

We employed the method described in Sect. 6.1. of \citet{Nogueras-Lara:2018aa} (the grid method), increasing the area of the region analysed by a factor of $\sim 10$. This method uses RC stars (giant stars in their helium core burning stage) \citep[e.g.][]{Girardi:2016fk} to compute simultaneously the extinction index and the extinction at a fixed wavelength ($\lambda = 1.61\ \mu m$). We used atmosphere models \citep{Kurucz:1993fk} to compute synthetic magnitudes of the RC stars for the filters used in our observations and minimised the corresponding $\chi^2$. We reddened the synthetic stellar models using a grid of extinctions and $\alpha$ (with a step of 0.016 for both of them). To model RC stars we used an effective temperature of 4750 K, log $g = +2.5$ \citep{2014ApJ...790..127B}, a radius of $10.0\pm0.5$ $R_\odot$ \citep[e.g.][]{Chaplin:2013kx,Girardi:2016fk}, and twice solar metallicity according to recent work \citep[e.g.][]{Do:2015ve,Feldmeier-Krause:2017kq, Nandakumar:2018aa,Do:2018aa,Schultheis:2019aa}, which allowed us to decrease the uncertainty of the results. We also assumed a distance to the GC of $8.0\pm0.1$ kpc with lower uncertainty, averaging the last results obtained by \citet{Gravity-Collaboration:2018aa} and \citet{Do:2019aa}. We selected the RC stars shown in the blue dashed parallelograms in Fig. \ref{CMD}. We expected some contamination of the red giant branch bump (RGBB)  \citep[see e.g.][]{Nataf:2011lq,Wegg:2013kx}, but since the intrinsic colour is similar to the RC \citep{Nogueras-Lara:2018ab} it does not have a significant influence on our results. We computed the extinction index and $A_{1.61}$ for all the RC stars detected in all three bands ($JHK_s$) with an uncertainty less than $0.05$ mag in all three bands ($\sim 62,000$ stars in total). To study the variability of the extinction index with the line of sight, we computed extinction index maps using the results obtained for $\alpha_{JH}$, $\alpha_{HK_s}$, and $\alpha_{JK_s}$. We defined a pixel size of 1 arcmin and computed the extinction index using a 3$\sigma$ clipping algorithm for all the RC stars within a pixel. We computed the maps for $\alpha_{JH}$, $\alpha_{HK_s}$, and $\alpha_{JK_s}$ to study the variation with  wavelength. We only calculated the extinction index value for a given pixel if more than 80 stars were detected. Figure \ref{maps} shows the obtained results for $\alpha_{JH}$, $\alpha_{HK_s}$, and $\alpha_{JK_s}$. We estimated the statistical uncertainties via the standard deviation of the distribution of the obtained values for each pixel. We found that the uncertainties are below $0.016$ for all the maps. The systematic uncertainties were not considered since they change all the pixels for each map in the same way, and we  analysed the relative difference between the pixels of the same map.

\begin{figure}
   \centering
   \includegraphics[width=\columnwidth]{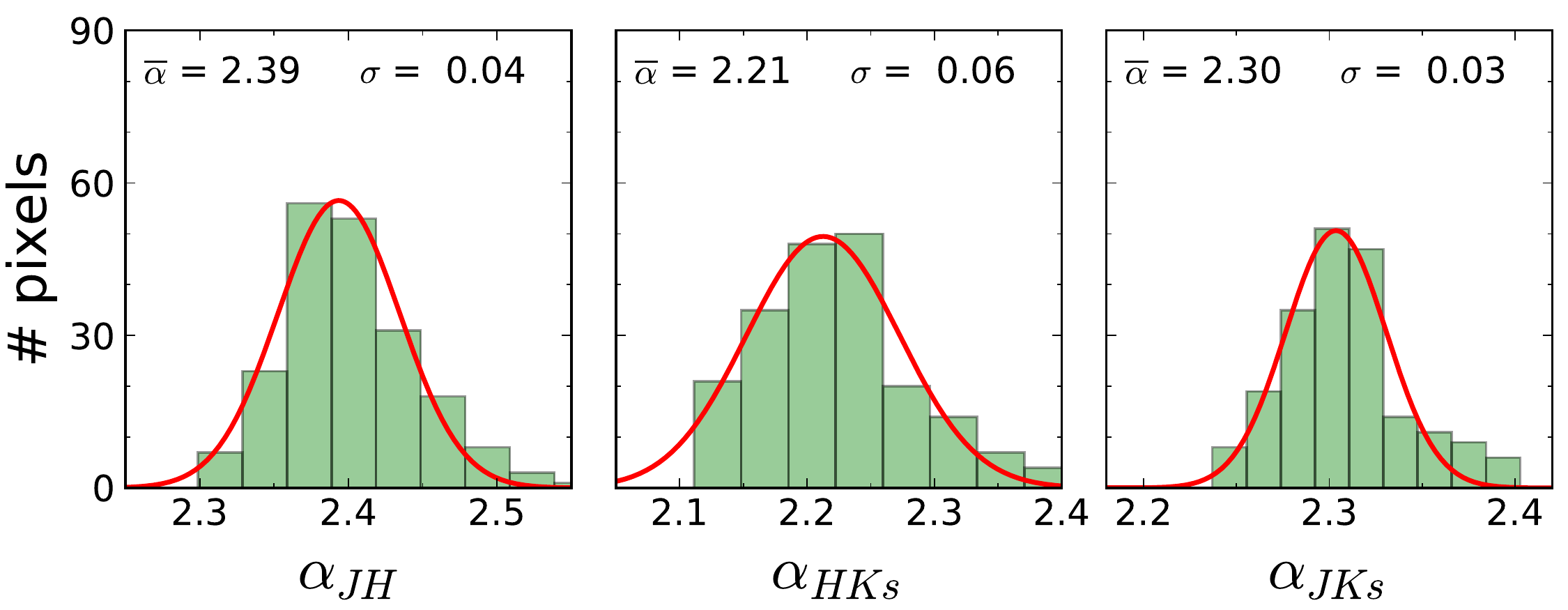}
   \caption{Histograms of the extinction-index values per pixel: a) $JH$-map; b) $HK_s$-map; c) $JK_s$-map. The red line indicates a Gaussian fit. The mean and the standard deviation are specified in the figure.}
   
   \label{varia}
    \end{figure}

We observed some variation between different pixels, as  can be seen in Fig. \ref{maps}. Nevertheless, we found that the systematic uncertainty of the ZP of the pointings used to produced the catalogue ($\sim 0.036$ mag in all three bands) can explain this variation. For this, we recomputed the extinction indices ($\alpha_{JH}$, $\alpha_{HK_s}$, and $\alpha_{JK_s}$) considering that, for each band independently, the magnitude of the RC stars used is affected by the systematic uncertainty of the ZP. We combined quadratically the obtained extinction index uncertainty for each band and found that the expected variation in the extinction index is $\Delta\alpha_{JH} = 0.05$, $\Delta\alpha_{HK_s} = 0.08$, and $\Delta\alpha_{JK_s} = 0.03$. We also created histograms of the extinction-index values per pixel. Figure \ref{varia} shows the results. The standard deviation of the distributions are below the ZP systematic uncertainties. We conclude that there is no significant variation in  the extinction index with the line of sight within the uncertainties of our data. This agrees with the non-variation in the extinction index measured between a region in the GC \citep{Nogueras-Lara:2018aa} and two regions in the inner bulge located at a distance of $\sim 0.4^\circ$ and $\sim 0.6^\circ$ (Galactic north) from Sgr\,A* \citep{Nogueras-Lara:2018ab}.

We also produced the corresponding maps for the extinction $A_{1.61}$ and checked that for all the band combinations ($JH$, $HK_s$, and $JK_s$), we obtained equivalent maps, as  was expected since we used  stars common to all three bands.

\subsection{Unique extinction index in the NIR?}

\begin{figure}
   \centering
   \includegraphics[width=\columnwidth]{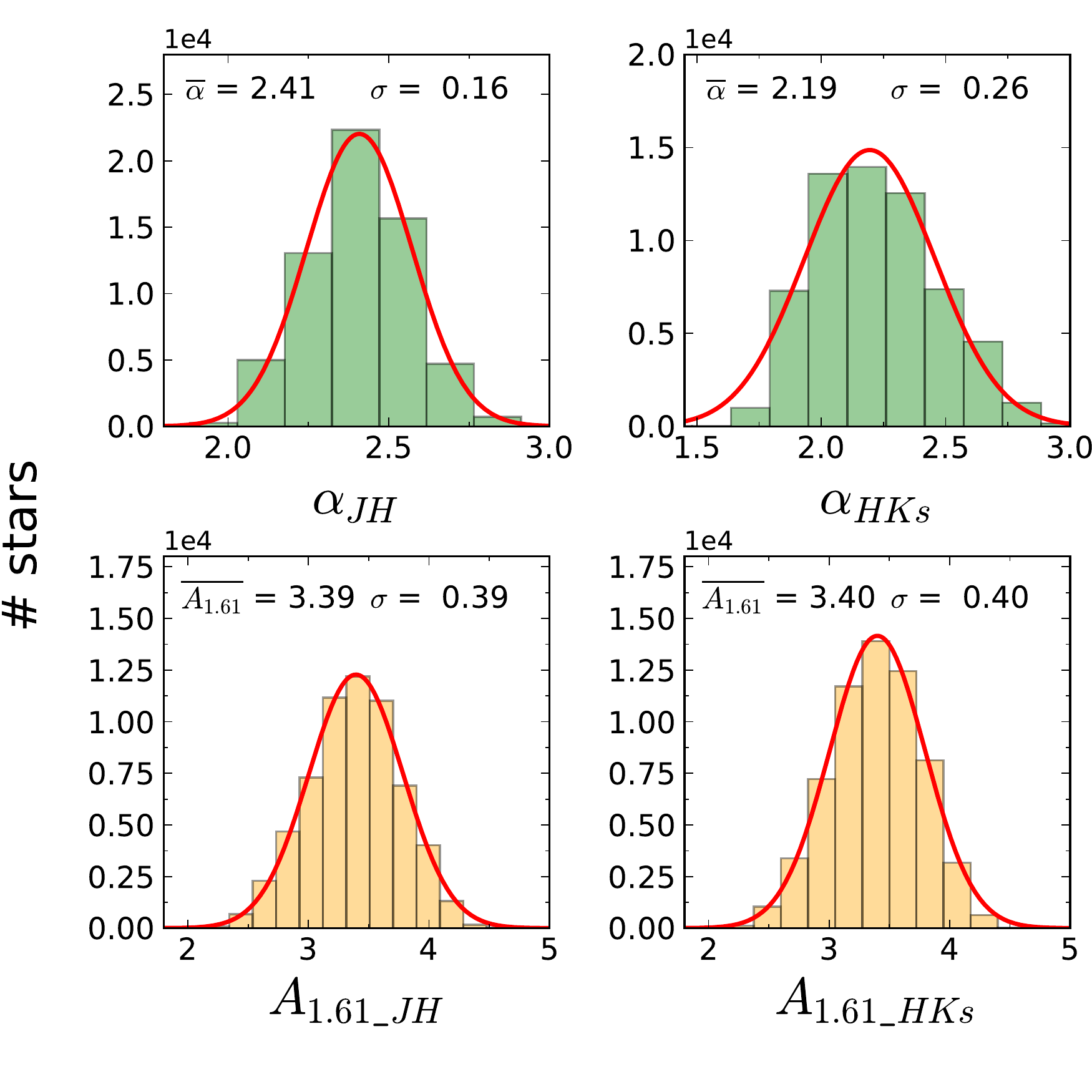}
   \caption{Upper panels: Histograms obtained for $\alpha_{JH}$ (left panel) and $\alpha_{HK_s}$ (right panel) using the method presented in Sect. \ref{grid_method}. Lower panels: Histograms obtained for $ A_{1.61}$, associated with the calculations using $JH$ and $HK_s$. The red lines show the Gaussian fits to the data. The mean and the standard deviation of each histogram are specified in each panel.}
   
   \label{hist_all}
    \end{figure}

To analyse the variation in the extinction index with  wavelength, we created histograms for the  values obtained for $\alpha_{JH}$, $\alpha_{HK_s}$, $\alpha_{JK_s}$, $\alpha_{JHK_s}$  and the corresponding extinctions $A_{1.61}$, for all the stars used in the analysis. The obtained distributions are well fitted  by a Gaussian model (Fig. \ref{hist_all}). Table \ref{alpha_grid} summarises the results. The uncertainties refer to systematics and were computed varying independently all the parameters involved in the calculation in their uncertainty ranges, as described in \citet{Nogueras-Lara:2018aa}. We  used a different range to estimate the uncertainties only  in the case of the distance to the GC and the metallicity of the GC stellar population, where the updated values allowed us to reduce the systematics in comparison to our previous work. The statistical uncertainties were estimated using the error of the mean of the distributions, and are negligible given the number of stars used for the calculation.

\begin{table}
\caption{Extinction index calculation following the method described in Sect. \ref{grid_method}.}
\label{alpha_grid} 
\begin{center}

\begin{tabular}{cccc}

 &  &  & \tabularnewline
\hline 
\hline 
 & Bands & $\alpha$ & $A_{1.61}$\tabularnewline
\hline 
 & $JH$ & $2.41$ $\pm$ 0.09 & $3.39$ $\pm$ 0.15\tabularnewline
Common & $HK_s$ & $2.19$ $\pm$ 0.14 & $3.40$ $\pm$ 0.14\tabularnewline
stars & $JK_s$ & 2.29 $\pm$ 0.09 & $3.47$ $\pm$ 0.18\tabularnewline
 & $JHK_s$ & $2.33$ $\pm$ 0.09 & $3.43$ $\pm$ 0.16\tabularnewline
\hline 
\end{tabular}

\end{center}
\vspace{0.5cm}
\textbf{Notes.} Only stars belonging to the RC detected in all three bands have been used.

 \end{table}

Our results suggest that the extinction index depends on wavelength in the NIR. We obtained $\Delta\alpha = \alpha_{JH}-\alpha_{HK_s} = 0.22\pm0.13$, which supposes a $\sim$ 2 $\sigma$ detection of a different extinction index between $JH$ and $HK_s$. We estimated the uncertainty computing the difference between the extinction indices when varying all the parameters specified in Sect. \ref{grid_method}. The uncertainty is lower than the value obtained simply using the quadratic propagation because the variation in some parameters produces a change in both $\alpha_{JH}$ and $\alpha_{HK_s}$ in the same direction.

\subsection{Variation in the extinction index with the extinction}

The spread of the RC along the reddening vector shown in Fig. \ref{CMD} is mainly due to differential extinction. In this way, we analysed the variation in $\alpha_\lambda$ with the extinction ($A_{1.61}$), dividing the RC stars in the CMD into small bins of $J-K_s=0.25$. We only used stars detected in all three bands with uncertainties $<0.05$ mag. The results are shown in Table \ref{alpha_ext}. The uncertainties are computed as explained in the Sect. \ref{grid_method}. We found some dependence of $\alpha_{HK_s}$ on the extinction, whereas $\alpha_{JH}$ appears to be constant. Nevertheless, we conclude that both extinction indices can be considered constant within the estimated uncertainties. On the other hand, we confirmed the previously computed value of $\alpha_{JH}-\alpha_{HK_s}\sim0.2$ that is also observed for different $A_{1.61}$.

%Since the $J$ band is more prone to extinction given its shorter wavelength, the number of stars detected in both $H$ and $K_s$ is larger. Therefore, we repeated the calculation for $\alpha_{HK_s}$ and $A_{1.61}$ using all the RC stars detected in both bands. In this way we increased around three times the number of stars used for the calculation. We obtained $\alpha_{HK_s} = 2.13 \pm 0.12$ and $A_{1.61} = 3.89$. Considering this $\alpha_{HK_s}$, we obtained an even larger difference $\Delta\alpha = \alpha_{JH}-\alpha_{HK_s} = 0.27\pm 0.11$. This supposes $\sim$ 3 $\sigma$ detection of a different extinction index between $JH$ and $HK_s$.
%

\begin{table}
\caption{Extinction index calculation for different extinctions (see Sect. \ref{grid_method}).}
\label{alpha_ext} 
\begin{center}

\def\arraystretch{1.4}

\small
\begin{tabular}{ccccc}

 &  &  &  & \tabularnewline
\hline 
\hline 
$J-K_s$ & $\alpha_{JH}$ & $\alpha_{HK_s}$ & $A_{JH}$ & $A_{HK_s}$\tabularnewline
\hline 
4-4.25 & 2.41 $\pm$ 0.10 & 2.14 $\pm$ 0.16 & 2.83 $\pm$ 0.15 & 2.83 $\pm$ 0.15\tabularnewline
4.25-4.5 & 2.40 $\pm$ 0.09 & 2.16 $\pm$ 0.15 & 3.04 $\pm$ 0.14 & 3.03 $\pm$ 0.14\tabularnewline
4.5-4.75 & 2.39 $\pm$ 0.09 & 2.19 $\pm$ 0.15 & 3.22 $\pm$ 0.15 & 3.22 $\pm$ 0.15\tabularnewline
4.75-5 & 2.39 $\pm$ 0.08 & 2.20 $\pm$ 0.14 & 3.44 $\pm$ 0.14 & 3.43 $\pm$ 0.14\tabularnewline
5-5.25 & 2.40 $\pm$ 0.08 & 2.22 $\pm$ 0.13 & 3.61 $\pm$ 0.16 & 3.61 $\pm$ 0.15\tabularnewline
5.25-5.5 & 2.40 $\pm$ 0.08 & 2.24 $\pm$ 0.13  & 3.78 $\pm$ 0.14 & 3.78 $\pm$ 0.15\tabularnewline
\hline 
 &  &  &  & \tabularnewline
\end{tabular}

\end{center}
\vspace{0.5cm}
\textbf{Notes.} Only stars belonging to the RC detected in all three bands have been used.

 \end{table}

\subsection{Slope of the RC features}

To check  the $\Delta\alpha = \alpha_{JH}-\alpha_{HK_s}$ obtained with the grid method, we studied the slopes of the RC features. We used all the RC stars shown in the blue dashed parallelograms in Fig. \ref{CMD}. Firstly, we divided the RC region in the CMD into small vertical bins to apply the SCIKIT-LEARN python function GaussianMixture \citep[GMM,][]{Pedregosa:2011aa} to compare a one-Gaussian model with a two-Gaussian model to fit the $K_s$ stellar distribution for each bin \citep{Nogueras-Lara:2018ab}. Using the Bayesian information criterion \citep{Schwarz:1978aa} and the Akaike information criterion \citep{Akaike:1974aa}, we found that a double-Gaussian model fits the data better as expected (\citet{Rui:2019aa}, Nogueras-Lara et al., submitted). We computed the slope of both RC features using a jackknife resampling method and calculated the systematic uncertainties varying the  bin width, the RC selection, and the width and the number of bins used, as described in \citet{Nogueras-Lara:2018ab}. We repeated the same analysis for the CMDs $K_s$ versus $J-K_s$ and $H$ versus $J-H$. 
The secondary RC feature is more sensitive to extinction and completeness problems given that it is fainter than the main feature. For this reason, we removed the last bins in the calculation of the slope of the secondary feature. Moreover, we excluded regions affected by dark clouds (using as reference the $J$ band, as it is more prone to extinction) that can influence the slopes of the features as they could change the relative number of stars in each feature for faint magnitudes. We also masked the central region belonging to the nuclear star cluster (NSC) because it could have a different star formation history and a lower completeness (Nogueras-Lara et al., submitted). Using the slope of the features, we computed the extinction index by means of Eq. 1 in \citet{Nogueras-Lara:2018ab}:

\begin{equation}
\label{eq_slope}
\alpha = -\frac{\log(1+\frac{1}{m})}{\log(\frac{\lambda_{\text{eff}_1}}{\lambda_{\text{eff}_2}})}.
\end{equation}

\noindent Here $m$ is the slope of the features in the CMD $\lambda_{\rm eff_2}$ versus $\lambda_{\rm eff_1}-\lambda_{\rm eff_2}$,  and $\lambda_{\rm eff_i}$ is the effective wavelength. Table \ref{alpha_slope} summarises the results obtained for each colour.

\begin{table}

\caption{Extinction index calculation using the slope of the RC.}
\label{alpha_slope} 

\begin{center}

\begin{tabular}{ccc}
 &  &  \tabularnewline
\hline 
\hline 
Bands &  & Extinction index  \tabularnewline
\hline 
$JH$ & $\alpha_1$ & $2.45 \pm 0.03 \pm 0.04$\tabularnewline
 & $\alpha_2$ & $2.59 \pm 0.04 \pm 0.06$ \tabularnewline
 \hline
 $HK_s$ & $\alpha_1$ & $2.26 \pm 0.01 \pm 0.01$ \tabularnewline
 & $\alpha_2$ & $2.32 \pm 0.02 \pm 0.01$ \tabularnewline
\hline 
 &  &   \tabularnewline
\end{tabular}
\end{center}

\vspace{0.1cm}
\textbf{Notes.} $\alpha_1$ is the extinction index found for the bright RC. $\alpha_2$ is the extinction index found for the faint RC. The uncertainties correspond to statistics and systematics, respectively. 

 \end{table}

Combining the values for both clumps, we computed $\alpha_{HK_s} = 2.29\pm0.02$ and $\alpha_{JH} = 2.52\pm0.09$, where the uncertainties were quadratically propagated. We obtained $\Delta \alpha = 0.23\pm0.09$. The uncertainty is even smaller if we just consider the bright RC, $\Delta \alpha = 0.19\pm0.05$, which supposes that the difference in the inter-band extinction index is detected with $\sim 4  \sigma$ significance. The results fully agree within the uncertainties with the previous values estimated using a completely independent method.

The extinction indices computed using this method are somewhat higher than those obtained in Sect. \ref{grid_method}. This could be  a consequence of a small shift, within the uncertainties, of the ZP calculation that affects the first method (which is considered in the estimation of the uncertainties) but does not affect the second method. Nevertheless, all the values agree within the uncertainties. Moreover, the ZP uncertainty does not affect the estimation of $\Delta \alpha$ in the second method. Therefore, we considered the value $\Delta \alpha = 0.19\pm0.05$ as the best estimation.

\section{Discussion and conclusion}

We have analysed the spatial variability of the extinction index and its dependence on the extinction, $A_{1.61}$. We find that there is no variation within the uncertainties. Therefore, it is possible to assume a constant $\alpha$ for the studied region of the NB in the NIR. We  detected a difference in the extinction index between $JH$ and $HK_s$ of $\Delta \alpha = 0.19\pm0.05$, combining the values obtained using two independent methods ($\Delta \alpha = 0.22\pm0.13$ and $\Delta \alpha = 0.19\pm0.05$). We used a weighted average for the calculation and the uncertainty estimation. We also obtain the mean values of the extinction indices $\alpha_{JH} = 2.43\pm0.03$ and $\alpha_{HK_s} = 2.23\pm0.03$, computed combining the results from Tables \ref{alpha_grid} and \ref{alpha_slope} (bright feature), and calculating the uncertainties via the difference between values (also coincident with the standard deviation). We did not use a weighted mean given the much lower uncertainties of the values obtained with the slope of the RC, which might lead to a biased result. Comparing our findings with those of  previous works, we suggest that some discrepancies towards different extinction-index values could be explained via  the direct assumption of having just one single $\alpha$ for $JHK_s$ \citep[e.g.][]{Nishiyama:2006tx,Stead:2009uq,Gosling:2009kl,Fritz:2011fk,Alonso-Garcia:2017aa}. This depends on the methodology used. In particular applying the method described in Sect. \ref{grid_method}, we obtained a value $\alpha_{JHK_s} = 2.32 \pm 0.09$, which lies between the two values computed for $\alpha_{JH}$ and $\alpha_{HK_s}$. On the other hand, using the slope of the RC to derive the extinction curve implies that it is necessary to know whether one or more RC features are present in the RC in the CMDs. Moreover, this method is quite dependent on the completeness of the photometry. The slope of the RC feature(s) might change at the faint end if the completeness is not sufficient.

To the best of our knowledge this is the first time that the extinction curve in the GC has been shown  not to depend on the line of sight or extinction, and to depend on wavelength. Previous studies always used a uniform extinction curve with different values. The great accuracy of this work has only been possible thanks to the high angular resolution GALACTICNUCLEUS survey.

  \begin{acknowledgements}
      The research leading to these results has received funding from
      the European Research Council under the European Union's Seventh
      Framework Programme (FP7/2007-2013) / ERC grant agreement
      n$^{\circ}$ [614922]. This work is based on observations made with ESO
      Telescopes at the La Silla Paranal Observatory under programme
      IDs 195.B-0283 and 091.B-0418. We thank the staff at
      ESO for their great efforts and helpfulness. F.N.-L. acknowledges financial support from a MECD pre-doctoral contract, code FPU14/01700. F.N. acknowledges financial support through Spanish grants ESP2015-65597-C4-1-R and ESP2017-86582-C4-1-R (MINECO/FEDER). We acknowledge support from the State Agency for Research of the Spanish MCIU through the Centre of Excellence Severo Ochoa Award of the Instituto de Astrof\'isica de Andaluc\'ia (CSIC) (SEV-2017-0709). N.N. acknowledges support from Sonderforschungsbereich SFB 881 `The Milky Way System' (subproject B8) of the German Research Foundation (DFG).
\end{acknowledgements}

\bibliography{/Users/fnoguer/Documents/Doctorado/My_papers/BibGC.bib}
\end{document}